\documentclass[12pt,manuscript,authoryear]{aastex}

\usepackage{graphicx}
\usepackage{natbib}
\usepackage{epstopdf}
\usepackage{grffile}

\begin{document}

\title{Migration of Gas Giant Planets in Gravitationally Unstable Disks}
  
\author{{\it Short Title: Gas Giant Migration ~~~~Article Type: Journal}}

\author{Scott Michael}
\affil{Department of Astronomy, Indiana University, Bloomington, IN 47405}
\email{scamicha@indiana.edu}

\and

\author{Richard H. Durisen}
\affil{Department of Astronomy, Indiana University, Bloomington, IN 47405}
\email{durisen@astro.indiana.edu}

\and 

\author{Aaron C. Boley}
\affil{Department of Astronomy University of Florida, Gainesville, FL 32611}
\email{aaron.boley@gmail.com}

\begin{abstract}
Characterization of migration in gravitationally unstable disks is necessary to understand the fate of protoplanets formed by disk instability. As part of a larger study, we are using a 3D radiative hydrodynamics code to investigate how an embedded gas giant planet interacts with a gas disk that undergoes gravitational instabilities (GIs). This Letter presents results from simulations with a Jupiter-mass planet placed in orbit at 25 AU within a 0.14 $M_{\odot}$ disk. The disk spans 5 to 40 AU around a 1 $M_{\odot}$ star and is initially marginally unstable. In one simulation, the planet is inserted prior to the eruption of GIs; in another, it is inserted only after the disk has settled into a quasi-steady GI-active state, where heating by GIs roughly balances radiative cooling. When the planet is present from the beginning, its own wake stimulates growth of a particular global mode with which it strongly interacts, and the planet plunges inward six AU in about 10$^3$ years. In both cases with embedded planets, there are times when the planet's radial motion is slow and varies in direction. At other times, when the planet appears to be interacting with strong spiral modes, migration both inward and outward can be relatively rapid, covering several AUs over hundreds of years. Migration in both cases appears to stall near the inner Lindblad resonance of a dominant low-order mode. Planet orbit eccentricities fluctuate rapidly between about 0.02 to 0.1 throughout the GI-active phases of the simulations. 
\end{abstract}


\label{firstpage}

\section{Introduction}

Migration of gas giant planets due to interactions with a circumstellar gas disk can play a major role in defining the architecture of planetary systems. Work on migration \citep[see review by][]{papaloizou2007} has included gravitational interaction of planets with both laminar and turbulent disks. However, radiative transport, detailed equations of state (EOS), and the self-gravity of the gas disk have usually been ignored; the effects of a non-isothermal EOS have only recently been included \citep[e.g.,][]{paardekooper2006,paardekooper2010,paardekooper2011}. Emerging studies, such as \citet{baruteau2011} and the one reported here, are beginning to address some of these issues. 
\citet{boss2005} and \citet{mayer2004} examined radial migration of planet-mass fragments in gravitationally unstable disks, but their disks were violently disrupted by fragmentation under conditions (radii $<$ 40 AU, disk-to-star mass ratios $M_d/M_s \sim 0.1$, and stellar mass $M_s \sim 1\;M_{\odot}$) where fragmentation may not actually occur  \citep{rafikov2005,rafikov2007,boley2006,boley2007b,boley2008,forgan2009,cai2010}. More recently, fragmentation into clumps with gas giant or brown dwarf masses has been documented in numerical simulations of disks that are relatively massive ($M_d/M_s \sim$ a few tenths) and spatially extended (outer radii $>$ 50 AU) \citep{krumholz2007,stamatellos2007,stamatellos2009,boley2009,boley2010}, where fragmentation is expected from semi-analytic arguments \citep[e.g.,][]{clarke2009,rafikov2009,dodson2009}. The fate of the clumps then depends in part on their radial migration, which is a chaotic and messy affair in a fragmenting disk \citep[e.g.,][]{boley2009,boley2010,vorobyov2010,boley2010b}. The occurrence of gravitational instabilities (GIs) may be episodic \citep[e.g.,][]{vorobyov2006,vorobyov2010b,zhu2010}. Clumps that survive and contract to the dimensions of young planets can later find themselves in a disk that erupts again into GI activity. As the star/disk system evolves, such a protoplanet may end up in a region of a GI-active disk where fragmentation does not occur.

To improve our understanding of how planets migrate in GI-active disks that are not fragmenting, we have begun a systematic study, using numerical 3D radiative hydrodynamics, where we investigate the effects on both the disk and the planet of inserting a planet-mass object into disks susceptible to GIs. Using techniques developed in earlier research by our group \citep{pickett2003, mejia2005,cai2006,cai2008,boley2006,boley2007b,michael2010}, we can identify the dominant spiral waves in a simulation and analyze how the waves interact with the planet's motion. Our goal is to determine both the effect of giant planets on GIs and the effect of GIs on planet migration. Because GIs are sensitive to radiative physics, we use a well-tested radiative scheme \citep{boley2007b} and realistic opacities \citep{dalessio2001}. 

Section 2 below presents our numerical methods and initial conditions.  We describe the simulation results in \S3, and discuss them in \S4.  

\section{Computational Methodology}

\subsection{3-D Radiative Hydrodynamics}

The CHYMERA (Computational HYdrodynamics with MultiplE Radiation Algorithms) code \citep{boley2007b} is a second-order, explicit, Eulerian scheme on a 3D cylindrical grid. The code uses a realistic equation of state for H$_2$ \citep{boley2007} and integrates an energy equation that includes $PdV$ work, net heating or cooling due to radiative flux divergence, and heating by artificial bulk viscosity. Calculations are done on a uniform cylindrical grid with reflection symmetry about the disk midplane and a grid size ($\varpi$,$\phi$,z) = (512,512,64). 
The $z$-axis is the rotation axis of the disk. The large number of azimuthal zones is necessary to resolve the planet's Hill sphere and the planet's wake. These simulations utilize the radiative cooling scheme developed and tested in \citet{boley2007b}, where the optically thick monochromatic flux in the $\varpi$ and $\phi$-directions is computed by flux-limited radiative diffusion and where the radiative transport of energy in the $z$-direction is solved using a one-ray discrete ordinate method in both optically thin and thick regions. Although the central star remains fixed at the grid center, we account for acceleration of the reference frame by the planet and by the disk via indirect potentials, as in \citet{michael2010}. The planet integration is done with a Verlet integrator \citep[e.g.,][]{hut1995}, and the indirect potential terms are treated as in \citet{nelson2000a}. The Rosseland mean and Planck mean opacities and molecular weights in our simulations are the same as those in \citet{boley2006,boley2007b}, except that we correct the \citet{dalessio2001} gas mean molecular weight to $\mu$ = 2.33.

\subsection{Initial Model and Simulation Conditions}

The model disk, based on an equlibrium model from \citet{pickett2003}, orbits a 1 $M_{\odot}$ star and has a mass $M_d = 0.14 M_{\odot}$, inner and outer radii at 5 and 40 AU, and an initial surface density distribution $\Sigma \sim \varpi^{-1/2}$. The time unit of one ORP (= outer rotation period) is defined as the rotation period of the initial disk at $\varpi \approx$ 32 AU, or about 180 yr. The disk is initially located between radial grid zones 30 and 240 and is close to isentropic, which results in a Toomre-$Q$ distribution with a marginally unstable \citep[see ][]{durisen2007} minimum value of 1.38 at radial grid zone 161 (26.7 AU). The computational grid extends radially to 512 zones to accommodate expansion of the outer disk once GIs become nonlinear. An outflow boundary condition is enforced at the upper vertical grid boundary, the outer radial grid boundary, and an inner radial boundary at 2 AU. To seed nonaxisymmetry, the density distribution is given an initial 0.01 \% random cell-to-cell perturbation.

This Letter presents three simulations. The first, which we call the {\sl fiducial run},  is simulated from time $t = 0$ to 21 ORP without a planet. In the second simulation, which we call the {\sl t = 0 planet run}, a one Jupiter-mass ($M_J$) planet with a roughly circular midplane orbit is included in the initial $t = 0$ ORP equilibrium disk at 25 AU. In the third simulation, which we call the {\sl t = 10 planet run}, a 1 $M_J$ planet with a similar orbit is inserted into the fiducial run at $t = 10.5$ ORP. These three simulations are part of a larger suite, to be described elsewhere, in which the planet mass is varied. We estimate semi-major axes $a$ and eccentricities $e$ from the planet's $\varpi(t)$ by using the maximum and minimum radii, $\varpi_{\rm max}$ and $\varpi_{\rm min}$, over each complete orbit of the planet, where one orbit is $\Delta\phi=2\pi$. Specifically, $a = (\varpi_{\rm max}+\varpi_{\rm min})/2$ and $e = (\varpi_{\rm max}-\varpi_{\rm min})/(\varpi_{\rm max}+\varpi_{\rm min})$. These parameters are intended to highlight the strong orbit-to-orbit variations in the motion of the planet.  A more refined estimate of the orbital elements is unnecessary because the potential of the disk, even if were axisymmetric, will cause departures from Keplerian dynamics.

\section{Results}

{\it The Fiducial Run.} As in \citet{mejia2005}, the fiducial run exhibits four distinct phases: initial {\sl axisymmetric cooling}, the onset of GIs in a well-defined {\sl burst} of a few discrete low-order modes, a {\sl transition} to more complex nonaxisymmetric structure, and the {\sl asymptotic} establishment of quasi-steady GI activity with an overall balance of heating and cooling. There is no fragmentation because this disk has a long cooling time \citep{gammie2001,boley2006,boley2007b}.

\begin{figure}[t]
\center
\includegraphics[width=11cm]{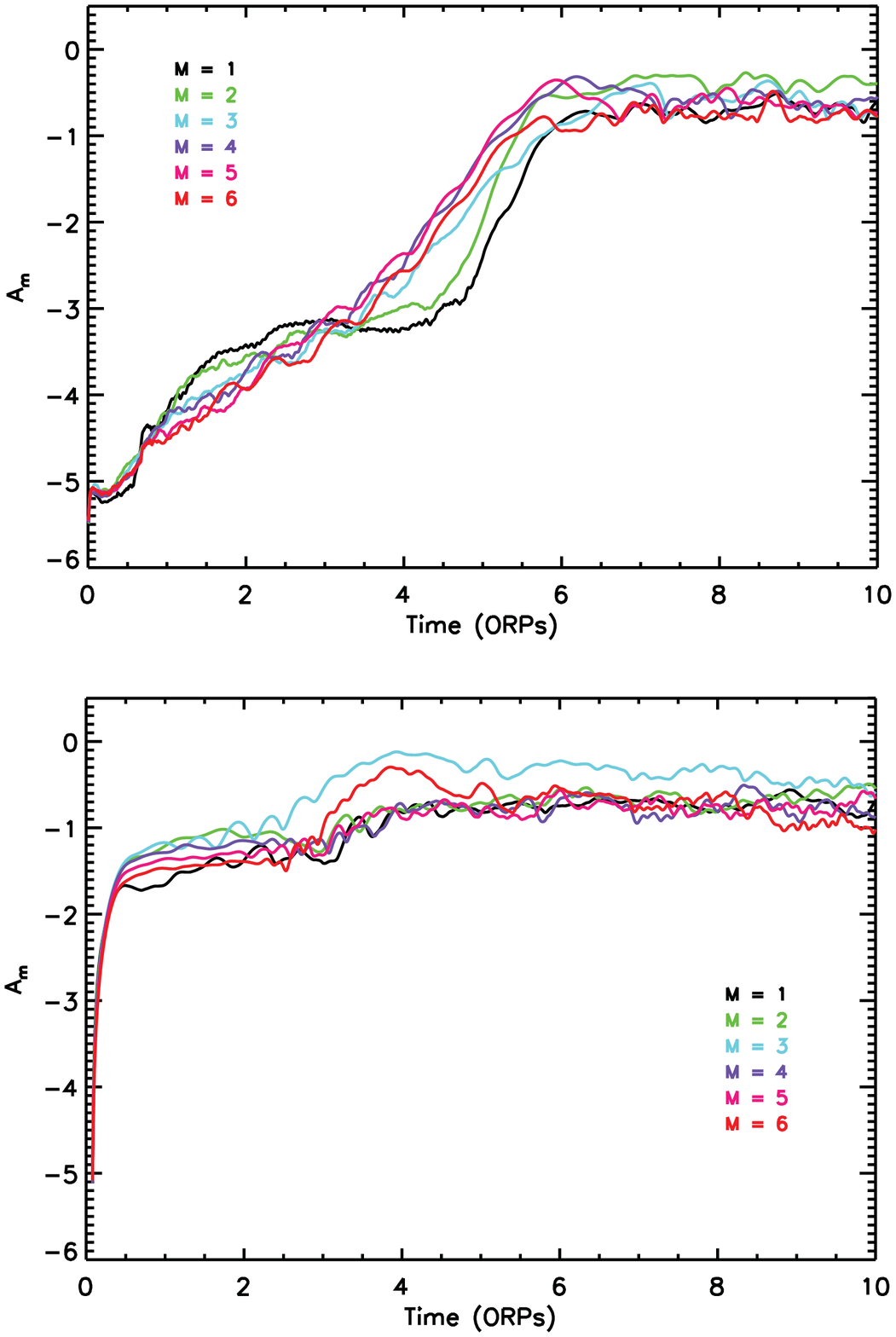}
\caption{(Color figure available in the online version) Global amplitudes of nonaxisymmetric density perturbations $A_m$ as a function of time for individual cos($m\phi$) perturbations. The formula for computing $A_m$ is equation (15) of \citep{boley2006}. Here the integrals are done only over the disk outside 15 AU to suppress contributions from edge effects in the inner disk. The top and bottom panels correspond to the fiducial run (no planet) and the $t = 0$ planet run, respectively.}
\label{fig:Am}
\end{figure}

{\it The $t = 0$ Planet Run.} Figure \ref{fig:Am} shows that inclusion of even this modest mass planet ($0.7 \%$ of $M_d$) has a dramatic effect on the burst phase. Without a planet, when all modes grow from imposed noise, it takes 4 ORPs for coherent spiral modes to organize and grow to nonlinear amplitude. The growth is centered near the $Q$-minimum at $26$ AU. Modes of cos($m\phi$) symmetry with $m = 4$ and 5 dominate. An $m = 3$ (three-armed) spiral also grows, but somewhat more slowly, and lags the other modes in time. In the $t = 0$ planet run, the planet develops an organized wake within about an ORP in which $m = 3$ is a dominant component ($\sim 5$ \% global density perturbation). This nonlinear seeding causes $m = 3$ to dominate the GI burst, which also occurs $\sim$ 2 ORP earlier. Because of our initial placement of the planet inside but close to the $Q$-minimum, the corotation radius (CR) of the $m = 3$ mode is fairly close to that of the planet's orbit radius. When the triggered $m = 3$ mode becomes strongly nonlinear at about 3 ORP, planet migration is significantly affected.

\begin{figure}[t]
\center
\includegraphics[width=12cm]{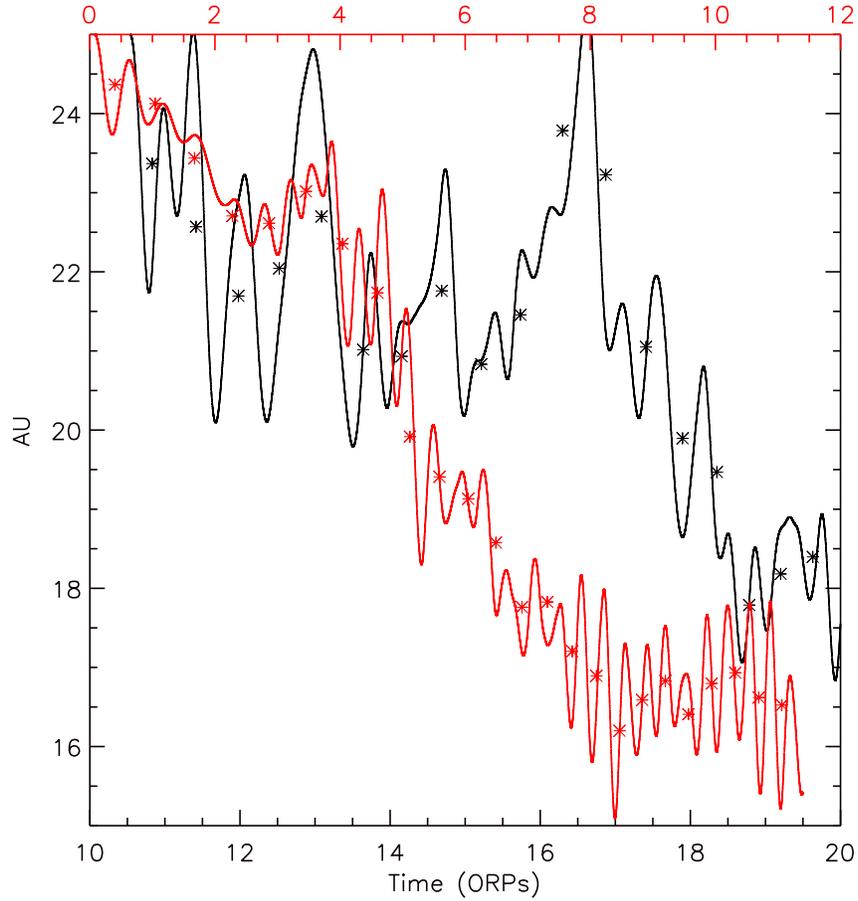}
\caption{Plot of the planet's radial position as a function of time $\varpi_p(t)$ for the two planet simulations. The dashed-diamond and solid-star lines (colored red and black in the online version) indicate the curves ($\varpi(t)$), points ($a$), and horizontal scales ($t$) that correspond to the $t = 0$ and $t = 10$ planet runs, respectively. The symbols are the approximate semi-major axes $a$ computed for each 0 to 2$\phi$ change in azimuth of the planet as explained in the text.}
\label{fig:a}
\end{figure}

Figure \ref{fig:a} shows the evolution of the planet's radial position. From 0 to 2.5 ORP, the planet is torqued primarily by its own wake and migrates inward. 
Beginning at about 3 ORP, the planet interacts with the now nonlinear $m = 3$ GI mode. At first, the planet gains angular momentum and moves outward, but, from 4 to 8 ORP, a time interval of  about 720 yr, the plant experiences a negative torque and plunges from 23 to 17 AU.
After $t = 8$ ORP, the main burst is over, and the disk transitions into its asymptotic state, where modes of many $m$-values become comparably strong (see Fig. \ref{fig:Am}). The planet's radial migration apparently stalls at about 16 to 17 AU. From an analysis of periodicities present in the gas disk between 8 and 12 ORP, the planet lies a few AU inside the inner Lindblad resonance of a strong $m = 2$ mode with CR at 29 AU in this early part of the asymptotic phase. 

{\it The $t = 10$ Planet Run.} The motion of the planet in this case is more difficult to interpret. We see intervals of fairly rapid inward or outward migration over several orbits, as well as times when radial migration appears to stall. Between $t = 15$ and 19 ORP, the pattern of outward migration for 2 ORP followed by an inward plunge over the next 2 ORP resembles the behavior in the $t = 0$ planet run during the burst, but, in this case, there is no distinct transition between phases of GI activity. Animations of the evolution of the midplane density (available at http://hdl.handle.net/2022/13304) show there is a complex interaction between the planet, its spiral wake, and the global spiral arms caused by the instabilities. Analysis of images and periodicities suggests that, between 15 and 19 ORP, the planet may be in a mean motion resonance with $m = 2$ and 3 patterns, both of which have CR at 27 AU. Over this time interval, the average orbit period of the planet is $\sim$ 0.50 ORP, while both the $m = 2$ and 3 spirals have a pattern period of $\sim 0.74$ ORP, suggesting a 3:2 resonance. Both of these modes are present, at nearly the same periodicities, in the fiducial run. So, in contrast with the $t = 0$ planet run, this planet's rapid phases of radial migration are caused by interaction with strong GI modes that exist independent of the planet.

\begin{figure}[t]
\center
\includegraphics[width=12cm]{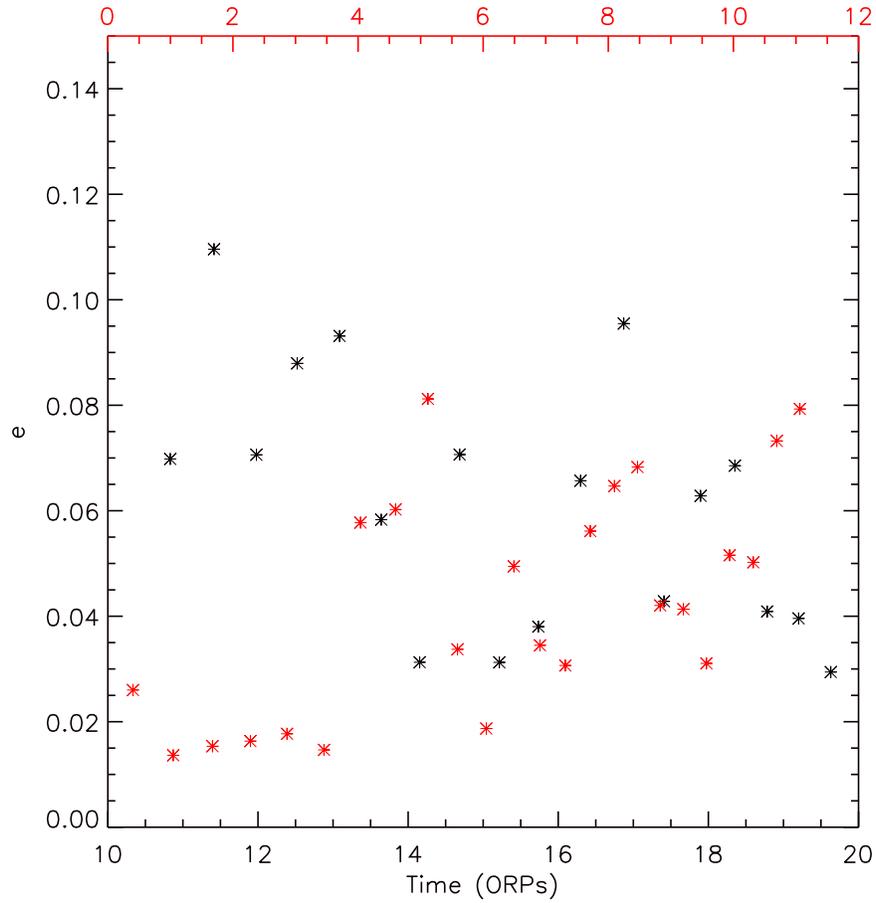}
\caption{Plot of the eccentricities $e$, computed for each 0 to 2$\phi$ change in azimuth of the planet as explained in the text, as a function of time. The diamonds and stars (colored red and black in the online version) indicate the points ($e$) and horizontal scales ($t$) that correspond to the $t = 0$ and $t = 10$ planet runs, respectively.}
\label{fig:e}
\end{figure}

{\it Eccentricity.} Figure \ref{fig:e} shows the evolution of $e$ in the planet simulations. In both cases, the planets were inserted with an approximate circular velocity computed by adding the interior disk mass to the stellar point mass. The presence of GI activity in the $t = 10$ planet run immediately increases $e$ to $\sim 0.08$. In the $t = 0$ planet run, the modest initial $e$ decreases during the 1.5 ORP when it is migrating only due to its own wake, as expected \citep{ward1998,ward2003,goldreich2004}. 
However, once there are strong nonlinear interactions with GI modes, $e$ jumps upward. In both runs, it appears that interaction with well-established GI activity leads to eccentricities between 0.02 and 0.1 that vary in a chaotic way between these extremes on orbit period time scales. The magnitude of $e$ is roughly consistent with the ratio of the sound speed to orbital speed of the disk and the modest values for the Mach numbers of the GI spirals \citep{boley2006a,boley2008}. In other words, the planet's orbit tends to have about as much nonaxisymmetry as reflected in the pitch angles of the GI-induced spirals \citep{cossins2009}.

{\it Migration Relative to the Disk.} The disk surface density distribution does not vary much in the asymptotic phase, so radial motion of the planet in the $t = 10$ planet run also represents radial migration relative to the gas. During the burst \citep[e.g.,][]{boley2006}, the surface density of the disk changes dramatically. To verify that the planet is really migrating relative to the background gas disk in the $t = 0$ run, we have compared the evolution of the planet's radial position $\varpi_p(t)$ with the mass inside that radius, expressed in terms of $m_{\rm cyl}$, the fraction of the disk mass interior to the planet. Between $t = 3.52$ and $3.89$ ORP, when $\varpi_p$ increases modestly, $m_{\rm cyl}$ increases from about 0.50 to 0.53; between 3.89 and 8.47 ORP, when $\varpi_p$ decreases dramatically, $m_{\rm cyl}$ also decreases dramatically from 0.53 to 0.30. So, the planet is migrating significantly with respect to the disk mass even while the inner part of the gas disk is moving inward during the burst. As a measure of the gas motion, $m_{\rm cyl}$ at 23 AU increases from 0.50 to 0.60 from $t =$ 3.89 to 8.47 ORP, corresponding to an average gas disk inflow rate of $\sim 10^{-5} M_{\odot}$ yr$^{-1}$.

\section{Discussion}

Even in non-self-gravitating disks, the presence of discrete radial structure is known to cause instabilities \citep{li2000,li2001}. \citet{meschiari2008} demonstrated that a gap, presumably opened by an embedded planet, can cause global modes to become unstable in self-gravitating disks that are otherwise stable against the development of GIs. Here we are dealing with a planet that is not quite massive enough to open a gap, but its gravitational interaction with the disk also has a strong effect on the growth and onset of GIs.
 
Figure \ref{fig:a} suggests that the overall radial drift during the first 10 ORP of the $t = 0$ planet run is not too different from what one would get by extrapolating the migration due solely to the planet's wake during the first 2 or 3 ORP, where the average radial migration is about 4$\times10^{-3}$ AU yr$^{-1}$. Although there are no simple formulae that are fully applicable to our radiatively cooled disk, using eq. (70) from \citet{tanaka2002} with parameters from the simulated disk, we estimate 6$\times10^{-3}$ AU yr$^{-1}$ as an analytic expectation. This is a reasonable estimate for comparison purposes, because our resolution is probably insufficient to compute the corotation torque \citep{paardekooper2010}. 
While the interaction with a GI-active disk does not seem to result in a drastically different overall migration rate, the direction and magnitude of migration in both simulations is quite variable and seems to depend strongly on interactions of the planet with discrete modes in the disk. In $t = 10$ planet run, the overall radial migration rate is similar but is even more erratic in direction than in the $t = 0$ planet run. The important departures from laminar disk theory are threefold: the more erratic nature of migration, the importance of interactions with discrete low-order modes, and stalling of migration near the inner edge of the strongest GI activity. The migration is neither a random walk nor monotonic but has a chaotic character due to nonlinear interactions. For all these reasons, it will be important in future studies to explore the dependence of the planet's behavior on its mass and on its placement relative to the site of GI-eruptions (for $t = 0$ type runs) and relative to the CRs of low-order GI modes (for $t = 10$ type runs).

In both simulations, the inward migration appears to halt near 17 AU. This may simply be a coincidence, because the overall average migration rates are similar, the starting positions are essentially the same, and the calculations have a similar duration. We plan to extend simulations like these to much longer times in the future. However, the final radial positions of both planets happen to be the location of the inner Lindblad resonance (ILR) for a strong m = 2 mode present in both runs. This ILR is associated with a surface density enhancement in the disk. Near this enhancement, the surface density gradient may be falling steeply enough to halt migration, a mechanism that can work in a laminar disk \citep{paardekooper2009}. 
Alternatively, at 17 AU, the level of GI activity is quite different interior to the planet ($Q \gtrsim 2$) and exterior to the planet ($Q \lesssim 2$), and this may affect the balance of the Lindblad torques exerted by the disk on the planet. Whether the migration really does stall and what the mechanism may be are subjects for further research.

{\it Summary.} The simulations in this Letter show that a planet of only 1$M_J$ placed near the $Q$-minimum in a 0.14$M_{\odot}$ disk can hasten the onset of gravitational instabilities and change the nature of the dominant modes during the initial burst. Although the magnitude of radial migration (a few$\times10^{-3}$ AU yr$^{-1}$) is not too different from that computed using the \citet{tanaka2002}  isothermal disk formula based on the Lindblad torques, it can fluctuate in sign on time scales on the order of the orbit period and can be strongly affected by gravitational interactions of the planet with discrete GI-driven spiral modes, leading to phases of rapid radial motion. Planet orbit eccentricities fluctuate between about 0.02 and 0.10. The simulations also suggest that planet migration may stall for long periods near the ILRs of dominant global GI-modes.

\section{Acknowledgments}

The authors would like to thank F.C. Adams and J.C.B. Papaloizou for useful encouragement in this project. The research was conducted with the support of NASA grants from the Origins of Solar Systems  Program (NNG05GN11G and NNX08AK36G). S.M. was supported during portions of this research by an NESSP Graduate Fellowship and by Indiana Space Grant Consortium Graduate Fellowships; A.C.B. is currently a Sagan Postdoctoral Fellow at U. Florida. This material is based upon work supported by the National Science Foundation under Grants No. ACI-0338618l, OCI-0451237, OCI-0535258, OCI-0504075, and CNS-0521433, and through TeraGrid resources provided by Indiana University, PSC, SDSC, and NCSA. TeraGrid systems are hosted by Indiana University, LONI, NCAR, NCSA, NICS, ORNL, PSC, Purdue University, SDSC, TACC, and UC/ANL. Any opinions, findings and conclusions or recommendations expressed in this material are those of the authors and do not necessarily reflect the views of the National Science Foundation.

\newpage

%\begin{figure}[t]
%\includegraphics{../Figures/name_of_figure.eps}
%\caption{here's the caption}
%\label{fig:name}
%\end{figure}

\label{lastpage}

\end{document}